\documentclass[pre,twocolumn,showpacs,showkeys,amsfonts,amssymb,amsmath,eqsecnum,letterpaper,nofootinbib]{revtex4}

\usepackage{graphicx}
\begin{document}
\renewcommand{\r}{\mathbf{r}}
\renewcommand{\k}{\mathbf{k}}
\newcommand{\K}{\mathbf{K}}
\newcommand{\exc}{\text{exc}}
\renewcommand{\DH}{\text{DH}}
\newcommand{\Ei}{\mathop{\mathrm{Ei}}}
\newcommand{\hx}{\tilde{x}}
\newcommand{\hX}{\tilde{X}}
\newcommand{\hW}{\tilde{W}}
\newcommand{\hPhi}{\tilde{\Phi}}
\newcommand{\hrho}{\tilde{\rho}}
\newcommand{\mynormalordering}[1]{::#1\-::\,}
\newcommand{\id}{\text{id}}

\newlength{\GraphicsWidth}
\setlength{\GraphicsWidth}{8cm}

\title{Description beyond the mean field approximation of an electrolyte
confined between two planar metallic electrodes}

\author{Gabriel T\'ellez}
\email{gtellez@uniandes.edu.co} 
\affiliation{Departamento de F\'{\i}sica, Universidad de Los Andes,
A.A.~4976, Bogot\'a, Colombia}

\begin{abstract}
We study an electrolyte confined in a slab of width $W$ composed of
two grounded metallic parallel electrodes. We develop a description of
this system in a low coupling regime beyond the mean field
(Poisson--Boltzmann) approximation. There are two ways to model the
metallic boundaries: as ideal conductors in which the electric
potential is zero and it does not fluctuate, or as good conductors in
which the average electric potential is zero but the thermal
fluctuations of the potential are not zero. This latter model is more
realistic. For the ideal conductor model we find that the disjoining
pressure is positive behaves as $1/W^3$ for large separations with a
prefactor that is universal, i.e.~independent of the microscopic
constitution of the system. For the good conductor boundaries the
disjoining pressure is negative and it has an exponential decay for
large $W$. We also compute the density and electric potential profiles
inside the electrolyte. These are the same in both models. If the
electrolyte is charge asymmetric we find that the system is not
locally neutral and that a non-zero potential difference builds up
between any electrode and the interior of the system although both
electrodes are grounded.
\end{abstract}



\maketitle

\section{Introduction}
\label{sec:Intro}

In this paper we study an electrolyte solution confined between two
parallel planar metallic electrodes. The study of the electrical
double layer near an electrode and more generally near any object
submerged in an electrolyte is of crucial importance in chemical
physics and in colloidal science. This problem was first considered by
Gouy~\cite{Gouy} and independently by Chapmann~\cite{Chapman} almost a
century ago. Their work is part of the foundations of colloidal
science~\cite{Verwey-Overbeek} and the physics of
electrolytes~\cite{McQuarrie}.

However their work and its developments are based on a mean field
description: the Poisson--Boltzmann equation. Although this mean field
approach describes accurately several properties of the systems, in
some situations it misses some subtle effects due to correlations. As
an example we can mention the old controversy about the possibility of
attraction between charged-like
colloids~\cite{Levine-Dube,Verwey-Overbeek} recently renewed by some
experimental results~\cite{Larsen-Grier,Kepler-Fraden,Carbajal-etal}.
It has been shown~\cite{Neu, Sader,
Trizac-Raimbault-charge-like-steric, Trizac-charge-like} that the mean
field approach (actually any local density approximation) cannot
predict any attractive effective interaction. Therefore the study of
electrolyte suspensions beyond the mean field approximation is
important.

This paper is oriented in that sense, although we will not consider
here the problem of charge-like attraction between colloids, but the
study of an electrolyte solution confined between two parallel
metallic planar electrodes beyond the mean field approximation. We
will be interested in questions like what is the force exerted on the
planar electrodes, it is attractive or repulsive, etc\dots? To have a
clear picture of the role of the correlations in this problem we will
consider the case when the two electrodes are grounded. The mean field
picture in this case is very simple: the mean field potential in the
electrolyte is zero everywhere and the fluid is uniform and locally
neutral. We will describe the first fluctuations around this mean
field picture in a low coupling regime where the average coulombic
energy of the microions of the solution is much smaller than their
thermal energy.

We should mention that this same problem was recently considered in
Ref.~\cite{Brandes-Lue}. However the authors of
Ref.~\cite{Brandes-Lue} made a mistake that has lead them to the wrong
conclusions, as explained below. The electrolyte is confined between
two conductor parallel planes. Each particle polarizes the
planes. There is an interaction energy between each particle and the
polarization charge that it induces in the electrodes. In
Ref.~\cite{Brandes-Lue} the authors forgot to include this energy in
the hamiltonian and this error make most of their conclusions
incorrect.

The outline of this paper and our main results can be summarized as
follows. In Sec.~\ref{sec:model-method} we present the models under
consideration and explain the
method~\cite{Torres-Tellez-finite-size-DH} used to find the
thermodynamic properties of the system. Actually there are two ways to
model the metallic electrodes. In a first model the boundaries are
supposed to be made of an ideal conductor material. This model is very
simple but it has the defect of neglecting the fluctuations of the
electric potential in the electrodes. In the second model, which is
more realistic, the electrodes are supposed to be genuine Coulomb
systems with very good conducting properties. We will call this model
the good conductor model. In this model the screening length vanishes
inside the conducting electrodes, the average electric potential is
zero in the electrodes, but the electric potential can fluctuate. We
will obtain results for both models and compare them using results
from Ref.~\cite{Janco-Tellez-ideal-conductor-limit}.

In Sec.~\ref{sec:pressure} we compute the grand potential of the
system and the pressure. For the ideal conductor model we find that
the disjoining pressure is positive. For large separations $W$ of the
slab, the disjoining pressure behaves as $1/W^3$. On the other hand,
for the good conductor model, the disjoining pressure turns out to be
negative and it decays as $e^{-2\kappa W}$ for large separations, with
$\kappa$ the inverse Debye length.

Finally in Sec.~\ref{sec:density} we find the microion density
profiles and the electric potential inside the electrolyte. These
quantities are the same regardless of the model used to describe the
electrodes (ideal conductor or good conductor). In that section we
retrieve the important
result~\cite{Aqua-Cornu-diel-wall1,Aqua-Cornu-diel-wall2,Aqua-these}
that for charge asymmetric electrolytes, a non-zero potential
difference builds up between each electrode and the middle of the
electrolyte solution and the system is not locally neutral although
both confining plates are grounded.

\section{Model}
\label{sec:model-method}

As explained in the Introduction, the system under consideration is an
electrolyte confined between two grounded conductor planar electrodes
separated by a distance $W$. Let us choose the $x$-axis in the
direction perpendicular to the electrodes, the origin is in the middle
of the electrodes and the electrodes are located at $x=\pm W/2$. We
will eventually also consider the limiting case when $W\to\infty$. In
this case we shall use the coordinate $X=x+W/2$ which measures the
distance from one electrode.

We will model the electrolyte as been composed of several species of
point-like microions with charges $q_{\alpha}$ labeled by a Greek
index. The solvent will be modeled as a continuum medium with
dielectric constant $\varepsilon$. As it is
well-known~\cite{LeboLieb-PRL, LeboLieb-AdvMat} a system of charged
point particles described by classical statistical mechanics is not
stable. In principle we should introduce some short-distance cutoff,
for instance the radius of the ions, to avoid the collapse of
particles of different sign. However in the low coupling regime
considered here most physical quantities have a well defined value
when this short-distance cutoff vanishes. Therefore our model will
give valuable quantitative information on the system provided the ion
radius is much smaller that the others lengths involved in the
problem.

The position of the $i$-th particle of the species $\alpha$
will be labeled as $\r_{\alpha,i}$. We shall work in the
grand-canonical ensemble at a reduced inverse temperature
$\beta=1/(k_B T)$, with $k_B$ the Boltzmann constant and $T$ the
absolute temperature. The average number of particles $\langle
N_{\alpha}\rangle$ of the species $\alpha$ is controlled by the
chemical potential $\mu_{\alpha}$. We shall use the fugacity
$\zeta_{\alpha}=e^{\beta u_{\alpha}}/\Lambda_{\alpha}^{3}$ where
$\Lambda_{\alpha}$ is the thermal de Broglie wavelength of the
particles which appears as usual in classical (i.e.~non-quantum)
statistical mechanics after the trivial Gaussian integration over the
kinetic part of the hamiltonian.  We shall impose the
pseudo-neutrality condition
\begin{equation}
\label{eq:pseudoneutral}
\sum_{\alpha}
q_{\alpha}\zeta_{\alpha}=0\,.  
\end{equation}
In the appendix B of Ref.~\cite{Torres-Tellez-finite-size-DH} it is
explained that this choice is equivalent to suppose that there is no
electric potential difference between the plates and the interior of
the system in the mean field approximation. The pseudo-neutrality
condition ensures the neutrality of the reservoir, however as we will
see below it does not ensure the global neutrality of the system
confined with Dirichlet boundary conditions. As we will show in
section~\ref{sec:density} there is a non-vanishing charge density near
the electrodes. However, this charge density induces in the electrodes
a polarization charge of opposite sign and the total charge (system
plus electrodes) is zero.

There are two models that can be used to describe the
electrodes~\cite{Janco-Tellez-ideal-conductor-limit,Alastuey-Janco-etc}. The
simpler one is the ideal conductor model. In this model, the
interaction potential between two unit charges located at $\r=(x,y,z)$
and $\r'=(x',y',z')$ is the solution of Poisson equation
\begin{equation}
  \Delta v(\r,\r')=-\frac{4\pi}{\varepsilon}\delta(\r-\r')
\end{equation}
satisfying the Dirichlet boundary conditions $v(\r,\r')=0$ if $x'=\pm
W/2$. It can be computed using, for example, the method of images,
\begin{widetext}
\begin{equation}
  \label{eq:Coulomb-potential-images}
  v(\r,\r')=\frac{1}{\varepsilon}\sum_{n=-\infty}^{+\infty}
  \left[
  \frac{1}{\left[(x-x'+2nW)^2+(\r_\perp-\r_\perp')^2\right]^{1/2}}
  -\frac{1}{\left[(x+x'+(2n+1)W)^2+(\r_\perp-\r_\perp')^2\right]^{1/2}}
  \right]
\end{equation}
\end{widetext}
with $\r_\perp=(y,z)$ the transversal part of the position vector $\r$
and $\varepsilon$ is the dielectric constant of the solvent. For
future reference we define the Coulomb potential for an unconfined
system
\begin{equation}
  \label{eq:Coulomb-v0}
  v^{0}(\r,\r')=\frac{1}{\varepsilon}\frac{1}{|\r-\r'|}
\end{equation}
which will be needed in the following. 

Notice that with this ideal conductor model, the microscopic electric
potential deduced from~(\ref{eq:Coulomb-potential-images}) vanishes on
the electrodes, for any configuration of the system. Then, not only
the average electric potential will vanish on the electrodes, but also
its fluctuations. 

In a real situation, even if the electrodes have very good conducting
properties, the electric potential inside the conductor will be
subjected to thermal fluctuations. This leads us to the other model
that can be used to describe the electrodes, the good conductor
model. In this model the electrodes are genuine Coulomb systems with
good conducting properties. We will restrict ourselves to a classical
(non-quantum) description of these conductors. We will consider that
the screening length in the conductors vanishes. This ensures that the
electrodes have good conducting properties. It is shown
in~\cite{Janco-Tellez-ideal-conductor-limit} that in this limit the
average electric potential vanishes on the electrodes. However the
fluctuations of the potential do not vanish in this limit.

Then a natural question arises: how are related the ideal conductor
model and the good conductor one? This question has already been
addressed in Ref.~\cite{Janco-Tellez-ideal-conductor-limit} and we
will make extensive use of the results presented there. It is shown in
Ref.~\cite{Janco-Tellez-ideal-conductor-limit} that the densities and
electric potential profiles inside the electrolyte are the same in
both models. Thus we will work with the (more simpler) ideal conductor
model for the determination of these profiles in
Sec.~\ref{sec:density}. 

On the other hand, the results for the grand potential and for the
pressure are different in both models, but there are simple relations
between them~\cite{Janco-Tellez-ideal-conductor-limit}. We will use
the ideal conductor model for the calculation of these quantities in
Sec.~\ref{sec:pressure} and using the relations found
in~\cite{Janco-Tellez-ideal-conductor-limit} we will obtain the
results for good conductor model. From now on we consider the ideal
conductor model unless stated otherwise.

Although to write down the hamiltonian of the system is a trivial
exercise in electrostatics, to clearly show what the problem is with
the previous study~\cite{Brandes-Lue} of this system, we will detail a
few (well-known) points before proceeding. First, consider the case
when only a planar electrode is located at $X=0$. Bringing first a
unit charge from infinity to a position $\r=(X,y,z)$ at a distance $X$
from the plane cost a non-zero energy, contrary to the case of an
unconfined system. This is because of the interaction between the
particle and the polarization charge it induces in the plane. In this
very simple geometry this interaction can also be understood as the
potential energy between the particle and an image charge located at
$\r^*=(-X,y,z)$. This energy is $-1/(4\varepsilon X)$ which can be
formally written as $(1/2)[v(\r,\r)-v^{0}(\r,\r)]$ (in this case
$v(\r,\r')$ is the potential
$(|\r-\r'|^{-1}-|\r-\r^{*'}|^{-1})/\varepsilon$ when only one
electrode is present). This interaction energy should be included in
the hamiltonian.

Following the same lines, in the general case of two metallic planes
the potential energy of the system reads
\begin{eqnarray}
H&=&\frac{1}{2}
\sum_{\alpha ,\gamma }\sum\nolimits_{i,j}^{^{\prime
}}q_{\alpha }q_{\gamma }v(\mathbf{r}_{\alpha,i},
\mathbf{r}_{\gamma,j})
\\
&+&
\frac{1}{2}\sum_{\alpha}\sum_{i}
q_{\alpha}^2 \left[
v(\mathbf{r}_{\alpha,i},\mathbf{r}_{\alpha,i})
-v^{0}(\mathbf{r}_{\alpha,i},\mathbf{r}_{\alpha,i})\right]
\,.
\nonumber
\end{eqnarray}
In the first sum the prime means that the case $\alpha=\gamma$ and
$i=j$ should be omitted. The second sum is the energy between each
particle and the polarization charge it has induced in the electrodes
as discussed previously. Introducing the microscopic charge density
defined as
\begin{equation}
\label{eq:rho}
\hat{\rho}(\mathbf{r})=\sum_{\alpha}\sum_{i} q_{\alpha }\delta
(\mathbf{r}-\mathbf{r}_{\alpha,i})
\end{equation}
we can formally write the potential part of the Hamiltonian of the
system as
\begin{eqnarray}
  \label{eq:H}
  H&=&\frac{1}{2}
  \int d\mathbf{r}\int d\mathbf{r}'\,\hat{\rho}(\mathbf{r})
  v(\mathbf{r},\mathbf{r}^{\prime })\hat{\rho}(\mathbf{r}^{\prime})
  \\
  &&-\frac{1}{2}\sum_{\alpha}\sum_{i=1}^{N_{\alpha }}q_{\alpha
  }^{2}v^{0}( \mathbf{r}_{\alpha,i},\mathbf{r}_{\alpha,i}) \,.
  \nonumber
\end{eqnarray}
The domain of integration in the first term is the space between the
two parallel electrodes ($-W/2<x<W/2$). Notice that from the first
term written in terms of ``continuous'' fields we subtract the
infinite ``self-energy'' of a particle $v^{0}(\r,\r)$ but with the
potential energy $v^{0}$ corresponding to an unconfined system.  

With the Coulomb potential $v^{0}$ given by Eq.~(\ref{eq:Coulomb-v0})
this self-energy is infinite. The appearance of infinite quantities
can be avoided and the subsequent analysis can be done more rigorously
by modifying the Coulomb potential $v^{0}(\r,\r')$ by introducing a
short-distance cutoff which smears out the singularity at
$\r=\r'$. Then, at the end of the calculations one can take the limit
of a vanishing cutoff. This has been done rigorously in
Ref.~\cite{Kennedy}. It turns out that the final results are the same
as if one takes the short-distance cutoff equal to zero from the
start. Therefore to simplify the algebra we will take from the start a
system of point particles without short-distance cutoff.

Now we follow the method proposed recently by the author and
collaborators in
Refs.~\cite{Torres-Tellez-finite-size-DH,Torres-Tellez-general-DH} to
study in general confined Coulomb systems in a low coupling
regime. Let us define the coulombic couplings $\Gamma_{\alpha}=\beta
q_{\alpha}^2 \zeta_{\alpha}^{1/3}/\varepsilon$. The method proposed in
Ref.~\cite{Torres-Tellez-finite-size-DH} is valid for
$\Gamma_{\alpha}\ll 1$.

In the method exposed in Ref.~\cite{Torres-Tellez-finite-size-DH} the
sine-Gordon transformation~\cite{Samuel} is performed in the
grand-canonical partition function, then the action of the
corresponding field theory is expanded to the quadratic order (valid
in the low coulombic coupling regime) around the stationary (mean
field) solution (here $\phi=0$). For details the reader is referred to
Ref.~\cite{Torres-Tellez-finite-size-DH} and to the
appendix~\ref{sec:appendix-A}. Then the grand partition function can
be written as
\begin{equation}
  \label{eq:Xi}
  \Xi=\frac{1}{Z_G}
  \int \mathcal{D}\phi 
  \, \exp[-S(\phi)]
\end{equation}
with
\begin{equation}
  \label{eq:Gaussian-integral}
  Z_{G}=\int \mathcal{D}\phi \, \exp\left[ -\frac{1}{2}\int
  \phi(\r) \left[-\frac{\beta\varepsilon \Delta}{4\pi}
  \right] \phi(\r)\,d\r\right]
  \,.
\end{equation}
and $S(\phi)$ is an action, quadratic in $\phi$, given by
\begin{equation}
  S(\phi)=
  \frac{1}{2}\int
  \frac{-\beta \varepsilon}{4\pi} \phi(\r)\Delta\phi(\r)
  +\sum_{\gamma}(\beta q_{\gamma})^2 \zeta_{\gamma} 
  \mynormalordering{\phi(\r)^2}\,d\r
\end{equation}
where $\mynormalordering{\phi(\r)^2}$ is a pseudo-normal ordered
product, see the appendix~\ref{sec:appendix-A} for details.

The field $\phi(\r)$ is a mathematical intermediary. At the mean field
level, the stationary equation for the action (before it is expanded
to the quadratic order) is Poisson--Boltzmann equation, and
$i\phi(\r)$ can be interpreted as the electric potential, however this
relation breaks down when we consider the fluctuations as in the
present case, for instance the correlations of $\phi(\r)$ are
short-ranged~\cite{Caillol-loop} whereas the correlations of the
electric potential are known to be
long-ranged~\cite{Lebo-Martin,Janco-screen-correl-revisit}. The
Gaussian functional integration in Eq.~(\ref{eq:Xi}) can be
performed~\cite{Torres-Tellez-finite-size-DH} to obtain
\begin{equation}
\label{eq:Xi-det}
\Xi=\left(
\prod_{n}\left( 1-\frac{\kappa^{2}}{\lambda _{n}}\right)
\prod_{m} e^{\frac{\kappa^{2}}{\lambda _{m}^{0}}} \right)
^{-1/2}e^{\sum_{\alpha }V\zeta _{\alpha }} 
\end{equation}
where $\lambda_n$ are the eigenvalues of the Laplacian operator
satisfying the Dirichlet boundary conditions and $\lambda_m^0$ are the
eigenvalues of the Laplacian operator defined in the whole space
$\mathbb{R}^3$ without boundaries. We will call this case in the
following the free boundary conditions case. The volume of the system
is $V$ and $\kappa=\sqrt{\sum_{\alpha} 4\pi \zeta_{\alpha} \beta
q_{\alpha}^2/\varepsilon}$ is the inverse Debye length. The second
product in Eq.~(\ref{eq:Xi-det}) involving $\lambda_{m}^{0}$ comes
from the subtraction of the self-energy term $v^{0}(\r,\r)$.

\section{Grand potential and pressure}
\label{sec:pressure}

\subsection{Ideal conductor model}
\label{sec:pressure-A}

In this part we use the ideal conductor model to describe the
electrode. As a remainder of this, we will use a superscript ``id'' in
the thermodynamic quantities.

\subsubsection{Grand potential}

For the present geometry the eigenvalues of the Laplacian for
Dirichlet boundary conditions and free boundary conditions
respectively are $\lambda=-\k^2-(n\pi)^2/W^2$ with
$n\in\mathbb{N}^{*}$ and $\k\in\mathbb{R}^2$ and
$\lambda_{m}^{0}=-\K^2$ with $\K\in\mathbb{R}^3$. We find that the
grand potential $\Omega^{\id}$ takes the form
$\Omega^{\id}=\Omega_{\text{ig}}+\Omega_{\exc}^{\id}$ with
$\Omega_{\text{ig}}=k_B T V \sum_{\alpha} \zeta_{\alpha}$ the ideal
gas contribution and $\Omega_{\exc}^{\id}$ the excess grand
potential. From Eq.~(\ref{eq:Xi-det}) we find the excess grand
potential $\omega_{\exc}^{\id}$ per unit area of a plate
\begin{widetext}
\begin{equation}
  \beta\omega_{\exc}^{\id}=
  \frac{1}{2(2\pi)^2} \int_{\mathbb{R}^2}
  \ln\prod_{n=1}^{\infty}\left(1+\frac{\kappa^2}{
    \left(\frac{n\pi}{W}\right)^2+\k^2}\right)\,d^2\k
  -\frac{W\kappa^2}{2(2\pi)^3} \int_{\mathbb{R}^3}\frac{d^3\K}{\K^2}
  \,.
\end{equation}
The product under the logarithm can be performed exactly~\cite{GR} to
obtain
\begin{equation}
  \beta\omega_{\exc}^{\id}=
  \frac{1}{4\pi} \int_{0}^{k_{\max}}
  \ln\left[
    \frac{k}{\sqrt{\kappa^2+k^2}}
    \frac{\sinh(W\sqrt{\kappa^2+k^2})}{\sinh(kW)}
    \right]
  \,k\,dk
  -\frac{W\kappa^2}{(2\pi)^2} \int_{0}^{K_{\max}} dK
  \,.
\end{equation}
\end{widetext}
Notice that we introduced two ultraviolet cutoffs $k_{\max}$ and
$K_{\max}$ for both integrals since each integral, taken separately,
is ultraviolet divergent. However together they should give a finite
result when $k_{\max}\to\infty$ and $K_{\max}\to\infty$ \textit{as far
as the bulk properties are concerned}. Indeed, in the limit
$W\to\infty$ we should recover the well-known bulk
result~\cite{Torres-Tellez-finite-size-DH, Kennedy} $\beta \omega_{b}
= -\kappa^3 W /(12\pi)$. This requirement imposes that the cutoffs
should be related by $K_{\max}=\pi k_{\max}/2$. Then doing the change
of variable $K=\pi k/2$ in the second integral the excess grand
potential per unit area can finally be written as
\begin{widetext}
\begin{equation}
  \label{eq:omega-exc}
  \beta\omega_{\exc}^{\id}
  =
  \frac{1}{4\pi}
  \int_{0}^{k_{\max}}
  \left[
    k\ln\left(
    \frac{k}{\sqrt{\kappa^2+k^2}}
    \frac{\sinh(W\sqrt{\kappa^2+k^2})}{\sinh(kW)}
    \right) -\frac{\kappa^2 W}{2}
    \right]\, dk\,.
\end{equation}
\end{widetext}
In principle we should take the limit $k_{\max}\to\infty$, however it
should be noted that the above expression has a logarithmic divergence
when $k_{\max}\to\infty$ which manifests itself in the surface
tension. This can be seen clearly if we expand $\omega_{\exc}$ for
$\kappa W\gg 1$,
\begin{equation}
  \label{eq:omega-large-W}
  \beta\omega_{\exc}^{\id}=-\frac{\kappa^3 W}{12\pi} + 2\beta\gamma +
  \frac{\zeta(3)}{16\pi W^2}+\mathcal{O}(e^{-2\kappa W})
\end{equation}
with the surface tension $\gamma$ given by
\begin{equation}
  \beta\gamma=\frac{\kappa^2}{16\pi}\left(\ln
  \frac{\kappa}{k_{\max}}-\frac{1}{2}\right)
\end{equation}
and $\zeta(3)$ is the Riemann zeta function evaluated at 3 (not to be
confused with the fugacities). In Eq.~(\ref{eq:omega-large-W}) all
terms that vanish when $k_{\max}\to\infty$ have been omitted. 

A few comments are in order. Concerning the surface tension $\gamma$
it is divergent when the cutoff $k_{\max}\to\infty$. This is normal:
it is due to the strong attraction that each particle and its images
of opposite charge in the electrodes feel. The small coupling regime
of an electrolyte near a plane metallic wall can also by studied from
a diagrammatic Mayer expansion. This is done in section 5 of
Ref.~\cite{Samaj-Janco-TCP-metal} for a two-dimensional Coulomb
system. These calculations can easily be adapted to a three
dimensional system to show that the surface tension $\gamma$ is
related to the integral of the screened interaction energy between a
particle and its image: $-\exp(-2\kappa X)/(4X)$. This energy is not
integrable at short distances and its integral has a logarithmic
divergence at $X=0$. In this picture one can impose a short-distance
cutoff $D$: the particles cannot approach below this distance to the
electrode, then the surface tension is proportional to $\ln \kappa
D$. Actually our ultraviolet cutoff $k_{\max}\propto 1/D$.

The second comment concerns the algebraic finite-size correction $k_B
T\zeta(3)/(16\pi W^2)$ to the grand potential.  This finite-size
correction is universal, it does not depend on the details of the
microscopic constitution of the system, and it has been proved to
exist even beyond the low coupling regime considered here provided
that the electrolyte is in a conducting phase and it has good
screening properties, in particular if it can screen an external
infinitesimal dipole~\cite{Janco-Tellez-coulcrit}. This correction is
a consequence only of the screening properties of the system, that
explains its universality. However as we will see later, this term is
only present for the ideal conductor model considered in this
section. For the good conductor model this algebraic correction in the
grand potential is absent.

We should also mention that this algebraic finite-size correction is
not present in the case of insulating
plates~\cite{Janco-Tellez-coulcrit, AttardMitchellNiham-JCP,
Dean-Horgan-two-loop}. Indeed if the boundary is made of a dielectric
material there is a subtle cancellation between the term found here
and the one from the Lifshitz interaction (the Casimir effect). For
dielectric boundary conditions the disjoining pressure has an
exponential decay $e^{-2\kappa W}$ at large separations $W$ and it is
attractive~\cite{AttardMitchellNiham-JCP, Dean-Horgan-two-loop} as
opposed to the present case of ideal conductor boundary
conditions. Further details on this interesting difference are
discussed in the appendix~\ref{sec:appendix-B}.


\subsubsection{Pressure}

%
%
\begin{figure}
\includegraphics[width=\GraphicsWidth]{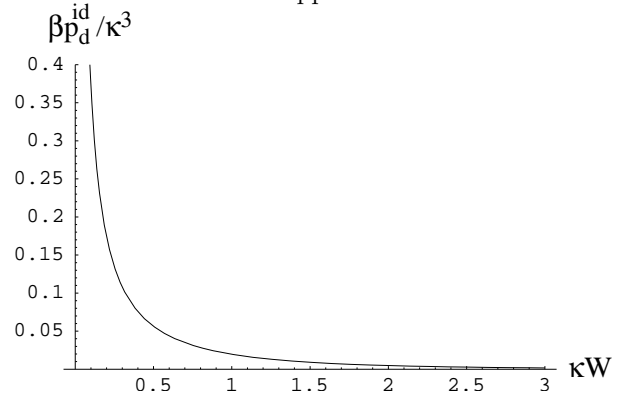}
\caption{
\label{fig:pressure-ideal-cond}
The disjoining pressure of the electrolyte confined by ideal conductor
electrodes. It is positive and always decreasing with increasing $W$
indicating that there is a repulsive force between the two ideal conductor
parallel plates. }
\end{figure}
%
%

The pressure is obtained from the usual relation
$p=-\partial\omega/\partial W$. From Eq.~(\ref{eq:omega-exc}) we find that
the excess pressure $p_{\exc}^{\id}$ is given by
\begin{widetext}
\begin{equation}
  \label{eq:p1}
  \beta p_{\exc}^{\id} = \frac{1}{4\pi} \int_0^{\infty} 
  \left[\frac{\kappa^2}{2}
  +k^2\coth(kW)-k\sqrt{k^2+\kappa^2}\coth(W\sqrt{k^2+\kappa^2})
  \right]\,dk\,.
\end{equation}
\end{widetext}
Although the grand potential has an ultraviolet divergence and should
be regularized as explained earlier, the pressure proves to be well
defined for $k_{\max}\to\infty$ (and $W\neq 0$). This is expected
since from the large-$W$ expansion~(\ref{eq:omega-large-W}) of the
grand potential we can see that the ultraviolet divergent part (the
surface tension contribution) does not depend on $W$. Notice however
that for $W\to 0$ the pressure is divergent. Let us mention that the
non-divergence of the pressure with the cutoff and more precisely the
fact that it is independent of the surface tension $\gamma$ is special
to this planar geometry. If we were to consider a confining geometry
with curved boundaries (for example an electrolyte confined in a
spherical domain) the surface tension would be a dominant term in the
pressure: due to the curvature $R$ the disjoining pressure for large
systems would be $p_d\propto-\gamma/R$, see
Ref.~\cite{MerchanTellez-jabon-anillos-tcp} for an example of this
effect.

Doing a few manipulations to Eq.~(\ref{eq:p1}) we can cast the
pressure in a form more adequate to study the disjoining pressure
$p_d^{\id}$, difference between the pressure $p^{\id}$ and the bulk
pressure $p^b$, and its large-$W$ behavior. The bulk pressure,
expressed in terms of the fugacities, is obtained from the limit
$W\to\infty$ of Eq.~(\ref{eq:omega-large-W}), and it is given by
\begin{equation}
  \label{eq:bulk-pressure}
  \beta p^b=\sum_{\alpha} \zeta_{\alpha}+ \frac{\kappa^3}{12\pi}
  \,.
\end{equation}
The well-known expression of the bulk pressure in terms of the
densities will be recovered in the next section,
Eq.~(\ref{eq:DH-equation-of-state}), when we obtain the expression of
the bulk densities in terms of the fugacities.

Then we find the disjoining pressure
\begin{widetext}
\begin{eqnarray}
  \label{eq:pdisj}
  \beta p_d^{\id}&=&\frac{\zeta(3)}{8\pi W^3} +
  \frac{\kappa^3}{4\pi}
  \int_0^{\infty}
  u\sqrt{u^2+1}
  \left[1-\coth\left(\kappa W\sqrt{u^2+1}\right)  \right]
  \, du
  \\
  &\underset{W\to\infty}{=}&
  \frac{\zeta(3)}{8\pi W^3} + \mathcal{O}(e^{-2\kappa W})
  \,.
\end{eqnarray}
\end{widetext}
For large separations $W$ of the electrode plates the disjoining
pressure is positive and decays as $1/W^3$. The force exerted on the
plates is therefore repulsive and the system is stable: if it is
compressed the pressure increases. This actually holds for any
separation as it can be seen in Fig.~\ref{fig:pressure-ideal-cond}.

\subsection{Good conductor model}

We now study the electrolyte confined by good conductor electrodes. We
will not use any superscript in the thermodynamic quantities in order
to differentiate them from the ones computed for the ideal conductor
model.

\subsubsection{Free energy and grand potential}

Let $F$ be the free energy of the system composed by the electrolyte
and the electrodes. These later ones are considered as Coulomb systems
with a vanishing screening length (high density of charged particles)
in the region $x<-W/2$ and $x>W/2$. The electrolyte is in the region
$-W/2<x<W/2$.

For the following discussion it is useful to introduce $F_0$ the free
energy of the electrodes in absence of the electrolyte. It is shown in
Ref.~\cite{Janco-Tellez-ideal-conductor-limit} that $F=F^{\id}+F_0$,
where $F^{\id}$ is the free energy of the electrolyte in the ideal
conductor model, which can be obtained from the results of the
previous section. The argument of
Ref.~\cite{Janco-Tellez-ideal-conductor-limit} can easily be applied
to the grand potential, thus
\begin{equation}
  \omega=\omega^{\id}+\omega_0
\end{equation}
with $\omega$ the grand potential (per unit area of the plates) of the
system when the electrodes are described as good conductors,
$\omega^{\id}$ the grand potential (per unit area) for the ideal
conductor model, which can be obtained from Eq.~(\ref{eq:omega-exc}),
and $\omega_0$ the grand potential of the good conducting electrodes
alone in the space in the absence of the electrolyte.

The large-$W$ expansion of this latter term $\omega_0$ is given by
Lifshitz theory~\cite{Lifshitz} in the classical regime, it is $-k_B T
\zeta(3)/(16\pi W^2)$. Thus it cancels the similar contribution that
we found in $\omega^{\id}$ in the last section,
Eq.~(\ref{eq:omega-large-W}). When the electrodes are described as
good conductors (which is a more realistic model) there is no
long-range contribution in $1/W^2$ to the grand potential of the
system as opposed to the ideal conductor model.

\subsubsection{Pressure}

The pressure is the force per unit area that the electrolyte exerts on
one plate, say the one at $x=W/2$. It can be computed by means of the
Maxwell stress tensor $T_{\mu\nu}$. It is $p=-T_{xx}$ evaluated at
$x=W/2$. In Ref.~\cite{Janco-Tellez-ideal-conductor-limit} it is shown
that the stress tensor in the ideal conductor model $T^{id}_{\mu\nu}$
and the one in the good conductor model $T_{\mu\nu}$ are related by
\begin{equation}
  \label{eq:diffTmunu}
  T_{\mu\nu}(\r)-T^{\id}_{\mu\nu}(\r)=
  \frac{\varepsilon}{4\pi}
  \left(
  \partial_\mu\partial_\nu'-\frac{\delta_{\mu\nu}}{2}\partial_{\sigma}
  \partial_{\sigma}'\right)
  G^{*}(\r,\r')\Bigg|_{\r'=\r}
\end{equation}
with $G^{*}(\r,\r')=v^{0}(\r,\r')-v(\r,\r')$ is (minus) the ``images''
contribution to the Coulomb potential $v$ in the ideal conductor
model.

\begin{figure}
\includegraphics[width=\GraphicsWidth]{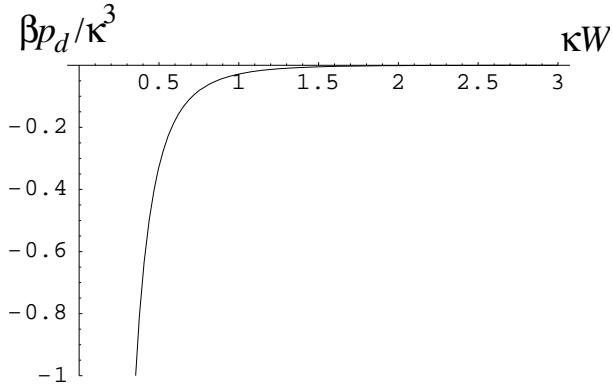}
\caption{
\label{fig:pressure-good-cond}
The disjoining pressure of the electrolyte confined by good conductor
electrodes. It is negative and always increasing with increasing $W$
indicating that there is an attractive force between the two conductor
parallel plates. }
\end{figure}

The r.h.s.~of Eq.~(\ref{eq:diffTmunu}) can be computed explicitly
giving
\begin{equation}
  T_{xx}-T^{\id}_{xx}=\frac{\zeta(3)}{8\pi W^3}
\,.
\end{equation}
Thus the pressure for the good conductor model is
\begin{equation}
  p=p^{\id}-\frac{\zeta(3)}{8\pi W^3}
\,.
\end{equation}
Using Eq.~(\ref{eq:pdisj}) we finally obtain the disjoining pressure
when the electrodes are modeled as good conductors
\begin{equation}
  \label{eq:pdisj-good-cond}
  \beta p_d=
  \frac{\kappa^3}{4\pi}
  \int_0^{\infty}
  u\sqrt{u^2+1}
  \left[1-\coth\left(\kappa W\sqrt{u^2+1}\right)  \right]
  \, du
  \,.
\end{equation}
Notice that since the function $\coth$ in the integrand is greater
than 1, the disjoining pressure is always
negative. Figure~\ref{fig:pressure-good-cond} shown a plot of the
disjoining pressure as a function of the width $W$. We notice that the
pressure is now an increasing function of $W$. This is just the
opposite behavior that the one obtained with the ideal conductor
model. Also for large-$W$ the disjoining pressure decays exponentially
as $e^{-2\kappa W}$.

%
%
%
%

\section{Density and electric potential profiles}
\label{sec:density}

The difference in the results for the pressure using the ideal
conductor model and the good conductor one are drastic. However the
results for the density and electric potential profiles inside the
electrolyte are the same in both models as it was shown in
Ref.~\cite{Janco-Tellez-ideal-conductor-limit}. Therefore we will
concentrate in this section on the ideal conductor model which is more
tractable.

\subsection{Density}

The density $n_{\alpha}(\r)$ can be obtained from the usual functional
derivative
\begin{equation}
\label{eq:func-deriv-density}
n_{\alpha}(\r)=\zeta_{\alpha}(\r)\frac{\delta \ln\Xi}{\delta
  \zeta_{\alpha}(\r)}\,.
\end{equation}
In the appendix~\ref{sec:appendix-A} it is shown that
\begin{widetext}
\begin{equation}
  \label{eq:density1}
  n_{\alpha}(\r)= \zeta_{\alpha} \left(
  1-\frac{\beta q_{\alpha}^2}{2} 
   \left[v_{\DH}(\r,\r)-v^{0}(\r,\r)\right]
  +\frac{\beta^2 q_{\alpha}}{2}
  \sum_{\gamma}\zeta_{\gamma}q_{\gamma}^3
  \int v_{\DH}(\r',\r)\left[v_{\DH}(\r',\r')-v^{0}(\r',\r')\right]
  \, d\r'\right)
\end{equation}
\end{widetext}
where $v_{\DH}(\r,\r')$ is the Debye--H\"uckel potential, solution of
Debye--H\"uckel equation
\begin{equation}
  \label{eq:DH-equation}
  \left(\Delta - \kappa^2\right)
  v_{\DH}(\r,\r')=-\frac{4\pi}{\varepsilon} \delta(\r-\r')
\end{equation}
satisfying the Dirichlet boundary conditions $v_{\DH}(\r,\r')=0$ if
$x'=\pm W/2$. Eq.~(\ref{eq:density1}) gives the density up to the
order $\Gamma_{\alpha}^{3/2}$ in the coulombic couplings. For the
present calculations we found that the most convenient form for
$v_{\DH}$ is as a Fourier transform in the transverse direction
$\r_{\perp}=(y,z)$. In Fourier transform, Debye--H\"uckel
equation~(\ref{eq:DH-equation}) reduces to an ordinary linear
differential equation in the $x$ variable, which can be easily
solved. Then we find
\begin{widetext}
\begin{equation}
  \label{eq:v_DH}
  v_{\DH}(\r,\r')=
  \frac{4\pi}{\varepsilon}
  \int_{\mathbb{R}^2}\frac{d\k}{(2\pi)^2}
  \frac{
    \sinh\left(\sqrt{\k^2+\kappa^2}\left(\frac{W}{2}-x'\right)\right)
    \sinh\left(\sqrt{\k^2+\kappa^2}\left(\frac{W}{2}+x\right)\right)
  }{
    \sqrt{\k^2+\kappa^2} \sinh\left(W\sqrt{\k^2+\kappa^2}\right)}
  \,e^{i\k\cdot\r_{\perp}}
\end{equation}
if $x<x'$ and exchange the roles of $x$ and $x'$ if $x'<x$.  Using
this expression into~(\ref{eq:density1}) we find that the density can
be expressed as
\begin{equation}
  \label{eq:density-res-intermed}
  n_{\alpha}(x)=\zeta_{\alpha}
  \left[
    1+\frac{\beta q_{\alpha}^2 \kappa}{2\varepsilon} f_{1}(\kappa x)
    +\frac{2\pi\beta^2 q_{\alpha}
    \sum_{\gamma}q_{\gamma}^3\zeta_{\gamma}}{\kappa\varepsilon^2}
    f_2(\kappa x)
    \right]
\end{equation}
with
\begin{equation}
  f_1(\hx)=-
  \int_0^{\infty}
  \left[
    \frac{2k}{{\sqrt{k^2 + 1}}}\,
    \frac{\sinh\left({\sqrt{k^2 + 1}}\,
      \left( \frac{\hW}{2} - \hx \right) \right)\,
      \sinh \left({\sqrt{k^2 + 1}}\,\left( \frac{\hW}{2} + \hx \right)
      \right)
    }{\sinh \left(\sqrt{k^2 + 1}\,\hW\right)} - 1
    \right]\,dk
\end{equation}
and
\begin{eqnarray}
  f_2(\hx)&=&
  \frac{\cosh\hx}{\cosh(\hW/2)}
  \int_0^{\infty}
  \left[1-
    \frac{4k\sqrt{k^2+1}}{4k^2+3}\coth \left(\hW\sqrt{k^2+1}\right)
    \right]\,dk
  \\
  &+&
  \int_0^{\infty}
  \frac{k\cosh\left(2\hx\sqrt{k^2+1}\right)
  }{\sqrt{k^2+1}\,(4k^2+3)\sinh\left(\hW\sqrt{k^2+1}\right)}
  \, dk
  \\
  &+&
  \int_0^{\infty}
  \left[
    \frac{k}{\sqrt{k^2+1}}\coth\left(\hW\sqrt{k^2+1}\right)-1
    \right] \, dk
\end{eqnarray}
where we have used distances measured in Debye length units
$\hx=\kappa x$ and $\hW=\kappa W$. Notice that the factors multiplying
$f_1$ and $f_2$ in the density, Eq. ~(\ref{eq:density-res-intermed}),
are of order $\Gamma_{\alpha}^{3/2}$ in the coulombic couplings. Our
approach neglects corrections of higher order than
$\Gamma_{\alpha}^{3/2}$.

After doing the change of variable $u=\sqrt{k^2+1}$ in the above
integrals some of them can be performed explicitly and doing some
manipulations we find the following convenient expressions for
$f_1(\hx)$ and $f_2(\hx)$
\begin{equation}
  f_1(\hx)=
  1+\frac{e^{-(\hW-2\hx)}}{\hW-2\hx}+\frac{e^{-(\hW+2\hx)}}{\hW+2\hx}
  + 2\int_1^{\infty}
  \frac{e^{-3u\hW}\cosh(2u\hx)\,du}{1-e^{-2u\hW}}+\frac{1}{\hW}\ln(1-e^{-2\hW})
\end{equation}
and
\begin{equation}
  f_2(\hx)=f_2^{(1)}(\hx)+f_2^{(2)}(\hx)-1-\frac{1}{\hW}\ln(1-e^{-2\hW})
\end{equation}
with
\begin{equation}
  f_2^{(1)}=
  \frac{\cosh\hx}{\cosh(\hW/2)}
  \left[
    1-\frac{\ln 3}{4}
    -8\int_1^{\infty}\frac{u^2\,e^{-2u\hW}}{(4u^2-1)(1-e^{-2u\hW})}\,du
    \right]
\end{equation}
and
\begin{subequations}
\begin{eqnarray}
  f_2^{(2)}(\hx)&=&\frac{1}{4}\left[
    e^{\frac{\hW}{2}-\hx}\Ei\left(-3\left(\frac{\hW}{2}-\hx\right)\right)
  -e^{-(\frac{\hW}{2}-\hx)}\Ei\left(-\left(\frac{\hW}{2}-\hx\right)\right)
    \right]
  \\&+& 
  \frac{1}{4}\left[
    e^{\frac{\hW}{2}+\hx}\Ei\left(-3\left(\frac{\hW}{2}+\hx\right)\right)
  -e^{-(\frac{\hW}{2}+\hx)}\Ei\left(-\left(\frac{\hW}{2}+\hx\right)\right)
    \right]
  \\
  &+&
  \label{eq:f22-expopetit}
  2\int_1^{\infty}
  \frac{e^{-3u\hW}\cosh(2u\hx)\,du}{(4u^2-1)(1-e^{-2u\hW})}
\end{eqnarray}
\end{subequations}
where $\Ei(z)=-\int_{-z}^{\infty} e^{-t}/t \,dt$ is the exponential
integral function. The advantage of these latter expressions is that
one can immediately see that the terms written as integrals are of
order $\mathcal{O}(e^{-2\hW})$ when $\hW\to\infty$. Therefore we can
easily obtain the expression for density in the case of one electrode
alone, with $X=x+W/2$,
\begin{multline}
  \label{eq:density-W-infty}
  n_{\alpha}(X)=\zeta_{\alpha} \Bigg[
    1+\frac{\beta q_{\alpha}^2 \kappa}{2\varepsilon}
    \left(1+\frac{e^{-2\kappa X}}{2\kappa X}\right)
    \\
    +
    \frac{2\pi\beta^2 q_{\alpha}
    \sum_{\gamma}q_{\gamma}^3\zeta_{\gamma}}{\varepsilon^2 \kappa }
    \left[
    e^{-\kappa X}\left(1-\frac{\ln 3}{4}\right)
    +\frac{e^{\kappa X} \Ei(-3\kappa X)
    -e^{-\kappa X} \Ei(-\kappa X)}{4}
    -1
    \right]
    \Bigg]
  \,.
\end{multline}
Far away from the metallic wall, $X\to\infty$, we find the bulk
density
\begin{equation}
  \label{eq:bulk-density}
  n_{\alpha}^b=\zeta_{\alpha} 
  \left( 1+ \frac{\beta q_{\alpha}^2 \kappa}{2\varepsilon}
  -\frac{2\pi\beta^2 q_{\alpha}
    \sum_{\gamma}q_{\gamma}^3\zeta_{\gamma}}{\varepsilon^2\kappa}
  \right)
  \,.
\end{equation}
Replacing back into Eq.~(\ref{eq:density-W-infty}) we find an
expression for the density profile in terms of the bulk density

\begin{subequations}  
\label{eq:density-W-infty-nbulk}
\begin{eqnarray}
  \label{eq:density-W-infty-nbulk-a}
  n_{\alpha}(X)=n_{\alpha}^b  \Bigg[&& 
    1+\frac{\beta q_{\alpha}^2 e^{-2\kappa_{\DH} X}}{4\varepsilon X}
    \\&&+
    \label{eq:density-W-infty-nbulk-b}
    \frac{2\pi\beta^2 q_{\alpha}
    \sum_{\gamma}q_{\gamma}^3 n_{\gamma}^b}{\varepsilon^2\kappa_{\DH}}
    \left[
    e^{-\kappa_{\DH} X}\left(1-\frac{\ln 3}{4}\right)
    +\frac{e^{\kappa_{\DH} X} \Ei(-3\kappa_{\DH} X)
    -e^{-\kappa_{\DH} X} \Ei(-\kappa_{\DH} X)}{4}
    \right]
    \Bigg]
  \nonumber\\
\end{eqnarray}
\end{subequations}
\end{widetext}
with corrections of higher order than $\Gamma_{\alpha}^{3/2}$. Here
$\kappa_{\DH}=\sqrt{4\pi\beta \sum_{\gamma} n_{\gamma}^b
q^2_{\gamma}/\varepsilon}$. We recover the expression that Aqua and
Cornu have previously obtained in their studies of the properties of a
classical Coulomb system near a
wall~\cite{Aqua-Cornu-diel-wall1,Aqua-Cornu-diel-wall2,Aqua-these}
using diagrammatic methods, up to a small difference: the first
term~(\ref{eq:density-W-infty-nbulk-a}) appears in~\cite{Aqua-these}
as an exponential Boltzmann factor of the screened interaction between
a particle and its image. Here, this exponential Boltzmann factors
appears linearized since the coulombic coupling is small. Our density
profile and the one found in~\cite{Aqua-these} will agree, in the low
coupling regime considered here, for distances not to close to the
electrode and our results are reliable in this case. On the other
hand, very close to the electrode, at distances comparable to the ion
radius, our results will differ from those of Ref.~\cite{Aqua-these},
this is a defect of the point-like model for the microions used here.

We can use Eq.~(\ref{eq:bulk-density}) which relates the fugacities to
the bulk densities into the expression~(\ref{eq:bulk-pressure}) of the
bulk pressure expressed in terms of the fugacities to recover the
well-known equation of state of Debye--H\"uckel
theory~\cite{McQuarrie}
\begin{equation}
  \label{eq:DH-equation-of-state}
  \beta p^b=\sum_{\alpha} n_{\alpha}^b - \frac{\kappa_{\DH}^3}{24\pi}
\,.
\end{equation}

Returning to the general case, for any arbitrary separation $W$ of the
plates it can be noticed that the density diverges at $x=\pm W/2$ as
$1/(x\mp W/2)$. The density does not have a finite value at the
contact of the electrodes but it diverges. This is a expected
behavior, since each particle is strongly attracted to its images in
the electrodes. This is related to the divergence of the surface
tension and the necessity to impose a short-distance minimum distance
of approach of the particles to the planar electrodes $D\propto
1/k_{\max}$ as explained in the previous section. The logarithmic
divergence in $\ln \kappa D$ of the surface tension is closely related
to the divergence of the densities as $1/(x\mp W/2)$ at the contact of
each electrode.

The charge density turns out to be 
\begin{equation}
  \rho(x)=\sum_{\alpha}q_{\alpha} n_{\alpha}(x)
  = \frac{\beta
    \kappa}{2\varepsilon}\left(\sum_{\alpha}\zeta_{\alpha} q_{\alpha}^3\right)
  \,\hrho(\kappa x)
\end{equation}
with the reduced charge density
\begin{widetext}
\begin{subequations}
\begin{eqnarray}
  \hrho(\hx)&=&f_1(\hx)+f_2(\hx)\\
  &=&
  \frac{e^{-(\hW-2\hx)}}{\hW-2\hx}+\frac{e^{-(\hW+2\hx)}}{\hW+2\hx}
  \nonumber\\
  &+&
  \frac{1}{4}\left[
    e^{\frac{\hW}{2}-\hx}\Ei\left(-3\left(\frac{\hW}{2}-\hx\right)\right)
  -e^{-(\frac{\hW}{2}-\hx)}\Ei\left(-\left(\frac{\hW}{2}-\hx\right)\right)
    \right]
    \nonumber\\
    &+& 
  \frac{1}{4}\left[
    e^{\frac{\hW}{2}+\hx}\Ei\left(-3\left(\frac{\hW}{2}+\hx\right)\right)
  -e^{-(\frac{\hW}{2}+\hx)}\Ei\left(-\left(\frac{\hW}{2}+\hx\right)\right)
    \right]
  \\
  &+&\left[1-\frac{\ln 3}{4}
    -8\int_1^{\infty}
    \frac{u^2 e^{-2u\hW}\,du}{(4u^2-1)(1-e^{-2u\hW})}
    \right]\frac{\cosh\hx}{\cosh(\hW/2)}
  \nonumber\\
  &+&
  8\int_1^{\infty}
  \frac{u^2 e^{-3u\hW} \cosh(2u\hx)\,du}{(4u^2-1)(1-e^{-2u\hW})}
  \,.
  \nonumber
\end{eqnarray}
\end{subequations}
\end{widetext}
In the case of a two-component symmetric electrolyte, $q_1=-q_2$, and
we have $\sum_{\gamma}\zeta_{\gamma}q_{\gamma}^3=0$, therefore the
system is locally neutral $\rho(x)=0$. For a general asymmetric
electrolyte $\sum_{\alpha} q_{\alpha}^3\zeta_{\alpha}\neq 0$ and the
system is not locally neutral. Furthermore the charge density diverges
near the plates as $1/(x\mp W/2)$ which is not integrable. Then the
total charge induced in the electrodes is infinite if the particles
are allowed to approach the electrodes as near as they can.

\begin{figure}
\includegraphics[width=\GraphicsWidth]{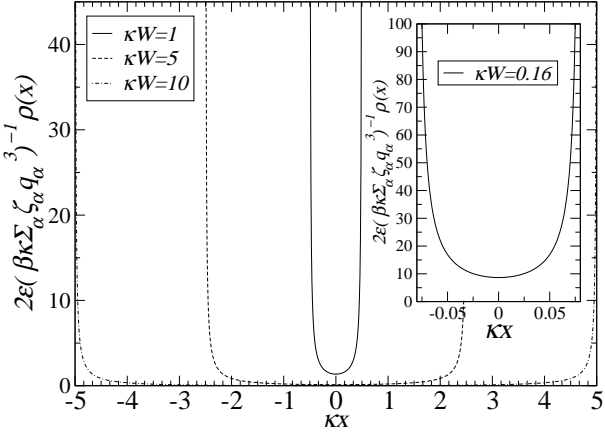}
\caption{
\label{fig:density-3plots}
The charge density profile in the slab for several values of the width
  $W$ at fixed $\kappa$.  }
\end{figure}

Fig.~\ref{fig:density-3plots} show several charge density profiles for
different values of $W$ with $\kappa$ fixed. As expected if $\kappa
W\gg 1$ the profiles for different values of $W$ are very similar
since the corrections to the case $W\to\infty$ are of order
$e^{-2\kappa W}$. This can be seen in the plots for $\kappa W=5$ and
$\kappa W=10$ in Fig.~\ref{fig:density-3plots}. The differences from
the case $W\to\infty$ can be only be noticed for small values of
$\kappa W$ as in the cases $\kappa W=1$ and $\kappa W=0.16$ of
Fig.~\ref{fig:density-3plots}. However let us remark that for any
value of $W$ the charge density from an electrode up to the middle of
the slab is strictly monotonous (increasing or decreasing depending on
the sign of $\sum_{\alpha} \zeta_{\alpha} q_{\alpha}^3$).

%
%
%
%


\subsection{Electric potential}

\begin{figure}
\includegraphics[width=\GraphicsWidth]{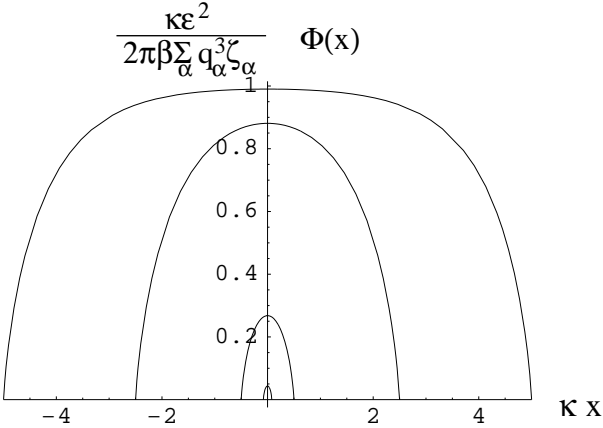}
\caption{
\label{fig:potential}
The electric potential profile $\Phi(x)$ for different values of the
width $W$ of the slab at fixed $\kappa$. From top to bottom $\kappa
W=10,$ 5, 1, 0.16.  }
\end{figure}

For the present geometry, the electric potential can be computed from
the charge density as
\begin{equation}
  \Phi(x)-\Phi(0)=\frac{4\pi}{\varepsilon}
  \int_0^{x} (x'-x)\rho(x')\,dx'
  \,.
\end{equation}
This gives
\begin{equation}
  \Phi(x)-\Phi(0)=
  \frac{2\pi\beta}{\varepsilon^2\kappa}
  \left[\sum_{\gamma}q_{\gamma}^3\zeta_{\gamma}  \right]
  \left(\hPhi(\kappa x)-\hPhi(0)\right)
\end{equation}
with the reduced electric potential
\begin{widetext}
\begin{subequations}
\label{eq:potential-ddp}
\begin{eqnarray}
  \hPhi(\hx)-\hPhi(0)&=&
  \frac{1}{2}\left[
    e^{\hW/2}\Ei(-3\hW/2)-e^{-\hW/2}\Ei(-\hW/2)
    \right]
  \\
  &+&
  \frac{1}{4}\left[
    e^{-(\frac{\hW}{2}+\hx)}
    \Ei\left(-\left(\frac{\hW}{2}+\hx\right)\right)
    -e^{\frac{\hW}{2}+\hx}
    \Ei\left(-3\left(\frac{\hW}{2}+\hx\right)\right)
    \right]
  \\
  \label{eq:terme-limite}
  &+&
  \frac{1}{4}\left[
    e^{-(\frac{\hW}{2}-\hx)}
    \Ei\left(-\left(\frac{\hW}{2}-\hx\right)\right)
    -e^{\frac{\hW}{2}-\hx}
    \Ei\left(-3\left(\frac{\hW}{2}-\hx\right)\right)
    \right]
  \\
  &+&
  \left[
  1-\frac{\ln 3}{4}-
  8\int_{1}^{\infty}
  \frac{u^2 e^{-2u\hW}\,du}{(4u^2-1)(1-e^{-2u\hW})}
  \right]
  \frac{1-\cosh \hx}{\cosh(\hW/2)}
  \\
  &-&
  2\int_1^{\infty}
  \frac{e^{-3u\hW}\left(\cosh(2u\hx)-1\right)}{(4u^2-1)(1-e^{-2u\hW})}\,du
  \,.
\end{eqnarray}
\end{subequations}
\end{widetext}
Fig.~\ref{fig:potential} shows the electric potential profile for
different values of the width $W$.

%
%
%
%

\begin{figure}
\includegraphics[width=\GraphicsWidth]{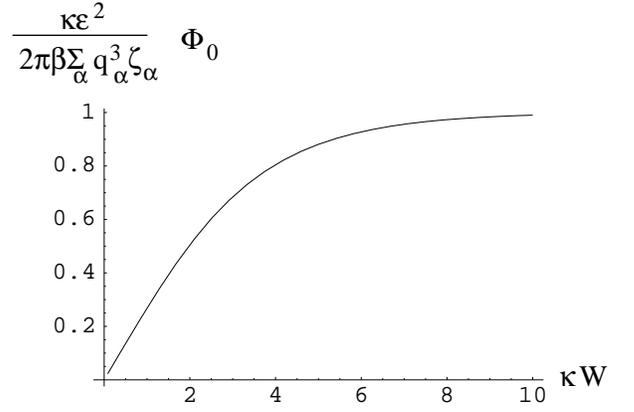}
\caption{
\label{fig:ddp-total}
The potential difference $\Phi_0$ between the middle of the slab and
  one electrode as a function of the width $W$ of the slab.  }
\end{figure}

An interesting quantity is the potential difference between a plate
(for example $x=W/2$) and the middle of the slab ($x=0$) which can be
obtained from the previous expression by replacing $x$ by $W/2$ (the
term~(\ref{eq:terme-limite}) in the previous equation has the limit
$-(\ln 3)/4$ when $x=W/2$). Fig.~\ref{fig:ddp-total} shows a plot of
the potential difference between the middle of the slab and a plate
$\Phi_0=\Phi(0)-\Phi(\pm W/2)=\Phi(0)$ as a function of $W$. It is
interesting to know the limit when $W\to\infty$.  From
Eq.~(\ref{eq:potential-ddp}) we get
\begin{equation}
\label{eq:ddp-total-max}
  \Phi_0
  \underset{W\to\infty}{=}
  \frac{2\pi\beta}{\varepsilon^2\kappa}
  \sum_{\gamma}q_{\gamma}^3\zeta_{\gamma}
  \,.
\end{equation}
For an asymmetric electrolyte a non-zero potential difference between
the middle of the electrolyte and any plate builds up although both
plates are grounded. The sign of this potential difference is given by
the parameter $\sum_{\alpha} \zeta_{\alpha} q_{\alpha}^3$. This
potential difference is a monotonous function (increasing or
decreasing depending on the sign of $\sum_{\alpha}q_{\alpha}^3
\zeta_{\alpha}$) of the width $W$ with an extremum value for
$W\to\infty$ given by Eq.~(\ref{eq:ddp-total-max}).

%
%
%
%

It is interesting to comment a few points on the case when only one
electrode is present, which has been previously studied by
Aqua~\cite{Aqua-these} using diagrammatic methods. In the limit
$W\to\infty$, with $\hX=\hx+\hW/2$, from Eq.~(\ref{eq:potential-ddp})
we recover Aqua's expression for the electric potential
\begin{eqnarray}
  \label{eq:potential-one-wall}
  \hPhi(\hX)-\hPhi_0&=&\left(\frac{\ln 3}{4}-1\right) e^{-\hX} 
  \\&+&
  \frac{1}{4}\left[ e^{-\hX} \Ei(-\hX)-e^{\hX} \Ei(-3\hX)\right]
  \,.
  \nonumber
\end{eqnarray}
We can notice that far away from the electrode the potential behaves as
\begin{eqnarray}
  \Phi(X)-\Phi_0& \underset{X\to\infty}{\sim}& 
  \frac{2\pi\beta}{\varepsilon^2\kappa}
  \left[\sum_{\gamma}q_{\gamma}^3\zeta_{\gamma}  \right]
  \left(\frac{\ln 3}{4}-1\right) e^{-\kappa X}
  \nonumber\\
  &=& \Phi_{\text{eff}}\ e^{-\kappa X}
\end{eqnarray}
where we defined
\begin{equation}
  \label{eq:effective-potential}
  \Phi_{\text{eff}}=  \frac{2\pi\beta}{\varepsilon^2\kappa}
  \left[\sum_{\gamma}q_{\gamma}^3\zeta_{\gamma}  \right]
  \left(\frac{\ln 3}{4}-1\right)
  \,.
\end{equation}
This result suggest the following interpretation. If we were to
understand this result using a mean field linearized
Poisson--Boltzmann equation, we can suppose that the electrode has an
effective potential $\Phi_{\text{eff}}$ given by
Eq.~(\ref{eq:effective-potential}). The potential of the electrode,
which is zero in our case, gets additively renormalized by
$\Phi_{\text{eff}}$ by the effect of the fluctuations around the mean
field. This interpretation follows the same philosophy that the one of
the theory of the renormalized charge in highly charged
colloids~\cite{Manning,Alexander}, except that in this case the
potential renormalization is due to the effect of the correlations and
not to the non-linear effects of the mean field theory. If the
electrode was at a fixed potential $V$, the effective potential as
seen far from the electrode would be
$V+\Phi_{\text{eff}}$~\cite{Aqua-these}.

In the spirit of this interpretation, notice that the renormalization
of the potential $\Phi_{\text{eff}}$ is positive if
$\sum_{\alpha}q_{\alpha}^3\zeta_{\alpha}$ is negative, and it is
negative otherwise. This potential renormalization only occurs for
asymmetric electrolytes. 

It is interesting to mention that a similar situation occurs in the
charge renormalization of colloids due to the non-linear effects in
the mean field approach for asymmetric electrolytes, although in the
other direction.  Indeed if the charge, say positive, of a colloid is
high enough to be in a non-linear regime, but small enough to be in a
non-saturation regime it has been found that the first deviation
(quadratic correction) of the effective charge from the bare charge
have the sign of
$\sum_{\alpha}q_{\alpha}^3\zeta_{\alpha}$~\cite{TellezTrizac-PB-asym,
Grahame, Ulander}. In particular in an intermediate regime the
effective charge of the colloid could be higher than the bare charge
if $\sum_{\alpha}q_{\alpha}^3\zeta_{\alpha}$ has the same sign as the
bare charge. This analogy is however only qualitative. The context in
the case of colloids is somehow different from the one considered
here. In the colloids the renormalization is due to the non-linear
effects of the mean field approach and here we considered the
corrections due to the correlations.


\section{Summary and conclusion}

We have obtained the first corrections due to fluctuations to the mean
field description of an electrolyte confined in a metallic slab of
width $W$. We considered two models to describe the metallic
electrodes. The ideal conductor model is more tractable, but it
neglects the fluctuations of the potential inside the electrodes. For
this (academic) model the disjoining pressure of the system is always
positive and it increases if the separation $W$ decreases indicating a
repulsive force on the metallic plates by the electrolyte and a stable
system. Also we confirmed~\cite{Janco-Tellez-coulcrit} that for large
separations $W$ the disjoining pressure has an algebraic decay in
$W^{-3}$, $p_d \sim [k_B T\zeta(3)/(8\pi)]\ W^{-3}$. This large-$W$
algebraic finite-size correction is universal: it does not depend on
the microscopic constitution of the system.

For the more realistic model of the good conductor electrodes, the
behavior of the pressure is completely different. There is no algebraic
decay in $1/W^3$ in the pressure. Its decay is exponential
$e^{-2\kappa W}$ at large distances. Furthermore the disjoining
pressure is negative, thus suggesting that there is an attractive
force between the electrodes.

We obtained some results for the density profiles and the electric
potential which are independent of the model used to describe the
electrodes. We retrieved~\cite{Aqua-Cornu-diel-wall1,
Aqua-Cornu-diel-wall2, Aqua-these} a very interesting behavior if the
electrolyte is asymmetric, in particular if $\sum_{\alpha}
q_{\alpha}^3 \zeta_{\alpha}\neq 0$. In this case the system is not
locally neutral, there is a local charge density with the same sign
that $\sum_{\alpha} q_{\alpha}^3 \zeta_{\alpha}$ near the
electrodes. Similarly the electric potential is not zero inside the
electrolyte although both plates are grounded: a potential difference
builds up between each electrode and the interior of the system. The
potential inside the electrolyte has the same sign that $\sum_{\alpha}
q_{\alpha}^3 \zeta_{\alpha}$.

As possible perspectives to this work let us mention the following.
Here we studied the first fluctuations corrections to the mean field
description of the electrolyte. We considered the most simple
situation at the mean field level: both plates are grounded and the
mean field potential is zero everywhere. We choose to study this
situation in order to show clearly the effects of
fluctuations. However our method can actually be extended to study
more general situations, for example when a potential difference is
imposed between the electrodes. In this case the mean field
description is the Poisson--Boltzmann theory studied by
Gouy~\cite{Gouy} and Chapman~\cite{Chapman} which can be a linear
theory if the potential difference is small or a non-linear theory if
the potential difference is high. On the top of this mean field
description, a generalization of our method can be used to find the
first corrections due to the microions correlations (whether the mean
field is linear or non-linear).

\begin{acknowledgments}
The author thanks B.~Jancovici for his comments on an earlier version
of the manuscript and for some discussions concerning the differences
between the ideal conductor and the good conductor models. I also
thank the referees for their comments, some of which led to the
development of appendix~\ref{sec:appendix-B} and to the comparison of
the ideal conductor and good conductor models. This work was partially
supported by COLCIENCIAS under project 1204-05-13625 and by
ECOS-Nord/COLCIENCIAS-ICETEX-ICFES.
\end{acknowledgments}


\begin{appendix}

\section{General expression for the densities}
\label{sec:appendix-A}

The density can be computed from the grand potential using
Eq.~(\ref{eq:func-deriv-density}). However to perform the functional
derivative for arbitrary fugacities $\zeta_{\alpha}(\r)$ we should
find a more general expression for the grand potential than the one
given by Eq.~(\ref{eq:Xi-det}) which is restricted to constant
fugacities satisfying the pseudo-neutrality
condition~(\ref{eq:pseudoneutral}). Similar calculations to the one
presented here can also be found in
Refs.~\cite{Caillol-low-fugas,Caillol-high-temp,Caillol-loop} in the
case of unconfined systems.

In general, the sine-Gordon transformation allows to write the grand
partition function without any approximation
as~\cite{Torres-Tellez-finite-size-DH,Samuel}
\begin{equation}
  \Xi=
  \frac{1}{Z_G} 
  \int \mathcal{D}\phi\,
  \exp[-S(\phi)]
\end{equation}
with $Z_{G}$ given by Eq.~(\ref{eq:Gaussian-integral}) and the action
$S$ given by
\begin{widetext}
\begin{equation}
  \label{eq:action}
  S(\phi)=-\int \left[
    \frac{\beta \varepsilon}{8\pi} \phi(\r) \Delta \phi(\r) + \sum_{\alpha}
    \zeta_{\alpha}(\r)e^{\beta q_{\alpha}^2 v^{0}(\r,\r)/2}
     e^{-i \beta q_{\alpha} \phi(\r)}\right]
  \,d\r
  \,.
\end{equation}
\end{widetext}
Let us define the Gaussian average 
\begin{equation}
  \langle\cdots\rangle_G=\frac{1}{Z_G}
  \int \mathcal{D}\phi \, 
  (\cdots)\,
  e^{ -\frac{1}{2}\int
    \phi(\r) \left[-\frac{\beta \varepsilon \Delta}{4\pi}
      \right] \phi(\r)\,d\r}
  \,.
\end{equation}
Notice that the covariance of the preceding functional Gaussian
measure is $\langle\phi(\r)\phi(\r')\rangle_G=\beta^{-1}
v(\r,\r')$. Therefore the last term of Eq.~(\ref{eq:action}) is very
similar to a normal ordering, since by definition
\begin{equation}
  \label{eq:true-normal-ordering}
  :\exp\left({-i \beta q_{\alpha}
  \phi(\r)}\right)\-:\,= e^{\beta q_{\alpha}^2 v(\r,\r)/2}
  e^{-i \beta q_{\alpha} \phi(\r)}
  \,.
\end{equation}
However the important difference is that in Eq.~(\ref{eq:action}) we
subtract the self-energy $v^{0}(\r,\r)$ for an unconfined system not
the self-energy $v(\r,\r)$ for a confined system. As previously
mentioned this has very important physical consequences for confined
systems. To proceed it is natural to define a pseudo-normal ordering as
\begin{equation}
  \label{eq:pseudo-normal-ordering}
  \mynormalordering{\exp\left({-i \beta q_{\alpha}
  \phi(\r)}\right)}= e^{\beta q_{\alpha}^2 v_{0}(\r,\r)/2}
  e^{-i \beta q_{\alpha} \phi(\r)}
\end{equation}
and write down the action as
\begin{equation}
  \label{eq:action-no}
  S=-\int \left[
    \frac{\beta\varepsilon}{8\pi} \phi(\r) \Delta \phi(\r) + \sum_{\alpha}
    \zeta_{\alpha}(\r)
    \mynormalordering{
     e^{-i \beta q_{\alpha} \phi(\r)}}\right]
  \,d\r\,.
\end{equation}
As we mentioned earlier in Sec.~\ref{sec:model-method} if we use the
Coulomb potential $v^{0}$ defined by Eq.~(\ref{eq:Coulomb-v0}) the
self-energy $v^{0}(\r,\r)$ is infinite. In principle one should
introduce a short-distance cutoff. Without this cutoff both quantities
$e^{-i\beta q_{\alpha}\phi(\r)}$ and $v^{0}(\r,\r)$ in
Eq.~(\ref{eq:pseudo-normal-ordering}) are not properly defined when
taken separately. However their combination in the
definition~(\ref{eq:pseudo-normal-ordering}) of the pseudo-normal
ordered exponential $\mynormalordering{\exp\left(-i \beta q_{\alpha}
\phi(\r)\right)}$ is well defined {\em even} in the case of a
vanishing short-distance cutoff. In particular later on we will
proceed to do an expansion of this exponential in powers of
$\mynormalordering{(-i\beta q_{\alpha}\phi(\r))^{n}}$. These
quantities are well defined for a vanishing short-distance cutoff and
they are of order $(\Gamma_{\alpha})^{n}$ in the coulombic coupling
constant.

For arbitrary position dependent fugacities the stationary point of
the action $S$ is $\phi=-i\psi$ with $\psi$ solution of the mean field
Poisson--Boltzmann equation
\begin{equation}\label{eq:PB}
  \Delta \psi(\mathbf{r}) + \frac{4\pi}{\varepsilon} \sum_{\alpha}
  \zeta_{\alpha}(\r) q_{\alpha} e^{-\beta q_{\alpha}
  \psi(\mathbf{r})} = 0 \,.
\end{equation}
Notice that if one takes constant fugacities satisfying the
pseudo-neutrality condition $\sum_{\alpha} \zeta_\alpha q_\alpha=0$
the solution of the mean field Poisson--Boltzmann equation is simply
$\psi(\r)=0$. 

Let us return to the general case of position dependent fugacities.
Expanding the action to the quadratic order in $\phi$ around the
stationary point leads to
$S(-i\psi+\phi)=S_{\text{mf}}+S_1+o(\phi^2)$, with
\begin{widetext}
\begin{equation}
  S_{\text{mf}}= 
  S(-i\psi)=
  \int\left[ \frac{\beta\varepsilon}{8\pi}
   \psi(\mathbf{r}) \Delta \psi(\mathbf{r})
   -\sum_{\alpha}\zeta_{\alpha}(\r) e^{-\beta q_{\alpha}
   \psi(\mathbf{r})} \right] \, d\mathbf{r}
\end{equation}
the action evaluated at the mean field solution and
\begin{equation}
 S_{1}=
  \frac{1}{2}\int
  \frac{-\beta\varepsilon}{4\pi} \phi(\mathbf{r})\Delta \phi(\mathbf{r})+
  \sum_{\alpha}
  (\beta q_{\alpha})^2 \zeta_{\alpha}(\r) e^{-\beta q_{\alpha}\psi(\mathbf{r})}
  \mynormalordering{\phi(\mathbf{r})^{2}}\, d\mathbf{r}  
  \,.
\end{equation}
\end{widetext}
We can now compute the functional
derivative~(\ref{eq:func-deriv-density}) with respect to the
fugacities to find
\begin{equation}
  \label{eq:density-funct-deriv-DH}
  n_{\alpha}(\r)=
  -\frac{\delta
    S_{\text{mf}}}{\delta\zeta_{\alpha}(\r)}
  -\frac{\int\mathcal{D}\phi\,\displaystyle
   \frac{\delta S_1}{\delta\zeta_{\alpha}(\r)}
    \,e^{-S_1}}{\int \mathcal{D}\phi\,e^{-S_1}}
  \,.
\end{equation}
However we should take special of the terms that depend on
the mean field $\psi(\r)$ since the latter is a function of the
fugacities via the Poisson--Boltzmann equation~(\ref{eq:PB}). In
particular from Eq.~(\ref{eq:PB}) we have
\begin{equation}
  \left(\Delta_{\r}-\kappa^2\right)
  \left.\frac{\delta\psi(\r)}{\delta\zeta_{\alpha}(\r')}\right|_0
  =-\frac{4\pi}{\varepsilon}\, q_{\alpha} \delta(\r-\r')
\end{equation}
where
$\left.\frac{\delta\psi(\r)}{\delta\zeta_{\alpha}(\r')}\right|_0$ is
evaluated for constant fugacities satisfying the pseudo-neutrality
condition~(\ref{eq:pseudoneutral}) and $\psi(\r)=0$. Then we can write
\begin{equation}
  \left.\frac{\delta\psi(\r)}{\delta\zeta_{\alpha}(\r')}\right|_0
  = q_{\alpha} v_{\DH}(\r,\r')
\end{equation}
with $v_{\DH}(\r,\r')$ the Debye--H\"uckel potential satisfying the
Debye--H\"uckel equation~(\ref{eq:DH-equation}) and the imposed
boundary conditions. Taking this into account we find the required
functional derivatives evaluated at constant fugacities satisfying
Eq.~(\ref{eq:pseudoneutral}) and $\psi(\r)=0$,
\begin{equation}
  \label{eq:derivada-Smf}
 \left. 
 \frac{\delta S_{\text{mf}}}{\delta \zeta_{\alpha}(\r)}
 \right|_0=-1
\end{equation}
and
\begin{widetext}
\begin{equation}
  \label{eq:derivada-S1}
  \left.
  \frac{\delta S_1}{\delta \zeta_{\alpha}(\r)}
  \right|_0=
  \frac{(\beta q_{\alpha})^2}{2} 
  \mynormalordering{\phi(\r)^2}
  -\frac{\beta^3 q_{\alpha}}{2}
  \sum_{\gamma}q_{\gamma}^3\zeta_{\gamma}
  \int v_{\DH}(\r',\r) 
  \mynormalordering{\phi(\r)^2}
  \,d\r
  \,.
\end{equation}
\end{widetext}
For constant fugacities the action $S_1$ reduces to
\begin{equation}
  \left.S_1\right|_0=
  \frac{1}{2}\int
  \frac{-\beta \varepsilon}{4\pi} \phi(\r)\Delta\phi(\r)
  +\sum_{\gamma}(\beta q_{\gamma})^2 \zeta_{\gamma} 
  \mynormalordering{\phi(\r)^2}\,d\r
  \,.
\end{equation}
If we define the average
\begin{equation}
  \langle\cdots\rangle_{\DH} =\frac{\int
  \mathcal{D}\phi\, (\cdots)\, e^{-\left.S_1\right|_0}}{\int
  \mathcal{D}\phi\,e^{-\left.S_1\right|_0}}
\end{equation}
we have
\begin{equation}
  \label{eq:covariance-DH}
  \beta \langle 
  \mynormalordering{ \phi(\r)^2 }
  \rangle_{\DH}=v_{\DH}(\r,\r)-v^{0}(\r,\r)
  \,.
\end{equation}
Then replacing~(\ref{eq:derivada-Smf}) and~(\ref{eq:derivada-S1}) into
Eq.~(\ref{eq:density-funct-deriv-DH}) and
using~(\ref{eq:covariance-DH}) gives Eq.~(\ref{eq:density1}) for the
densities.


\section{On the importance of the boundary conditions}
\label{sec:appendix-B}

As it was mentioned in section~\ref{sec:pressure} there is an
importance difference on the behavior of the disjoining pressure in
the case with ideal conductor boundaries and the case when the
electrolyte is confined by a walls made of material with a dielectric
constant $\varepsilon_w$~\cite{AttardMitchellNiham-JCP}. Namely, for
the latter case the pressure has an exponential decay $e^{-2\kappa W}$
at large separations $W$ whereas in the case presented here we found
an algebraic decay in $1/W^3$ which is furthermore universal,
i.e.~independent of the microscopic detail: notice that the
coefficient of $W^{-3}$ in Eq.~(\ref{eq:pdisj}) is just a number
independent of the Debye length $\kappa^{-1}$ and of the other
microscopic parameters.

Although the ideal boundary conditions case considered here is
formally obtained when $\varepsilon_w=\infty$, this limit has a very
different behavior than in the case $0<\varepsilon_w<\infty$ (the
ideal dielectric boundaries case, $\varepsilon_w=0$, is also special,
in that case there is also an algebraic decay at large separations
$W^{-3}$ in the pressure~\cite{Janco-Samaj-TCP-dielec} similar to the
one found here). This difference is not only present on the behavior
of the pressure but also in the correlation functions. For instance,
it is known that for dielectric boundaries with
$0<\varepsilon_w<\infty$ the charge correlation along the walls have
an algebraic decay as $|\mathbf{r}_{\perp}|^{-3}$ where
$\mathbf{r}_{\perp}$ is the direction parallel to the
walls~\cite{Janco-plane-wall-II, Aqua-Cornu-dipolar-PRE} whereas for
ideal conductor boundaries $\varepsilon_w=\infty$ (or ideal dielectric
$\varepsilon_w=0$) this decay is faster that any power law (see the
review Ref.~\cite{Martin-sum-rules-review} and references cited
therein).

In this appendix we show how both kinds of boundary conditions can be
related and understand the presence of the universal term in $1/W^{3}$
in the Dirichlet boundary conditions case and its absence in the case
of insulating boundaries. The following analysis relies on a
macroscopic description of the electrolyte in terms of collective
modes which actually disregards the microscopic detail of the system
but can give a correct description of some universal properties of the
system for instance the presence of the $1/W^{3}$ term in the
pressure.

Let us consider an electrolyte confined in the slab domain $D$ with
separation $W$ and with boundaries made of a dielectric material with
dielectric constant $0<\varepsilon_w<\infty$. For simplicity and
without lost of generality (divide $\varepsilon_w$ by $\varepsilon$)
we will take the dielectric constant of the solvent
$\varepsilon=1$. It is well-known that the electric potential $\Phi$
of a linear collective mode of oscillation with frequency $\omega$ of
this electrolyte can be described a Laplace type of equation
\begin{equation}
\chi_\omega  \Delta \Phi=0
\end{equation}
inside the domain $D$, with an effective dielectric constant
$\chi_\omega=1-\omega_p^2/\omega^{2}$ and $\omega_p$ is the plasma
frequency (see for instance~\cite{Tellez-modes}). Outside the domain
where the electrolyte is confined $\Phi$ satisfies a Laplace equation
$\Delta\Phi=0$ since there are no real charges outside. The potential
should also satisfy the boundary conditions:
$\Phi_{\text{in}}=\Phi_{\text{out}}$ and
$\varepsilon_w\partial_n\Phi_{\text{out}}=
\chi_{\omega}\partial_n\Phi_{\text{in}}$, where $\partial_n\Phi$
denotes the component of (minus) the electric field normal to the
boundary.

One can distinguish between two type of modes. If $\chi_{\omega}\neq
0$, $\Phi$ has a vanishing Laplacian inside the domain: there are no
charges inside the domain. Only at the boundaries there are some
surface charges. These are the surface modes. They represent a system
of two parallel walls in the vacuum with fluctuating surface
charges. Then the contribution of the surface modes to the pressure is
the same as the one of the Lifshitz theory~\cite{Lifshitz} in the
classical (i.e.~non-quantum) limit~\cite{Forrester-Janco-Tellez,
Levin-charge-attract}. This contribution comes from the well-known
Casimir forces, it is attractive and for large separations it is given
by $-k_B T \zeta(3)/(8\pi W^{3})$.

If $\omega=\omega_p$, $\chi_{\omega}=0$ and charges inside the domain
can exist since $\Delta\Phi\neq 0$ inside $D$ is acceptable. These are
the volume modes which oscillate all at the same frequency
$\omega=\omega_p$, the plasma frequency. Due to the boundary
conditions, the potential for volume modes satisfy Neumann boundary
conditions outside the domain $\partial_n \Phi_{\text{out}}=0$, and
since $\Phi$ is harmonic outside and vanishes at infinity, this
implies that $\Phi=0$ everywhere outside the domain $D$. In conclusion
the volume modes represent a system of volume charges that satisfy
Dirichlet boundary conditions for the potential. The volume modes of
the electrolyte confined by dielectric boundaries are very similar to
the system we studied here with ideal conductor boundaries. If one
computes the contribution to the pressure coming from these volume
modes one will find the same result that the one we have found in
section~\ref{sec:pressure-A}, namely that the pressure has an
universal algebraic decay for large separations $W$ given by $+k_B T
\zeta(3)/(8\pi W^{3})$. This contribution gives a repulsive force and
is exactly the opposite as the one coming from the surface modes.

Adding both contributions from the surface and the volume modes one
finds that the algebraic contributions to the pressure cancel each
other. In conclusion for the system confined by dielectric boundaries
there is no term in $1/W^{3}$ in the pressure as previously
noted~\cite{AttardMitchellNiham-JCP}. However for ideal conductor
boundary conditions, a repulsive term in $1/W^{3}$ for the pressure is
present.

The above analysis is somehow similar to the one done in the
comparison between ideal conductor and good conductor model, in the
sense that the absence of the $1/W^3$ algebraic term in the pressure
in the dielectric boundaries case and the good conductor case is due
to a cancellation between the Lifshitz term and the one from the ideal
conductor model. However the specific details are different.

\end{appendix}



\begin{thebibliography}{99}
\bibitem{Gouy} G.~Gouy, J.~de Physique (France) \textbf{IX}, 457
  (1910).

\bibitem{Chapman}
D.~L.~Chapman, Phil.~Mag.~\textbf{25}, 475 (1913).

\bibitem{Verwey-Overbeek} E.~J.~W.~Verwey and J.~Th.~G.~Overbeek,
  \textit{Theory of the stability of lyophobic colloids}, Elsevier
  Publishing Company (New York, 1948), Dover Publications (New York,
  1999).

\bibitem{McQuarrie} see, e.g., D.~A.~McQuarrie, \textit{Statistical
Mechanics}, Harper Collins Publishers, New York (1976).

\bibitem{Levine-Dube} S.~Levine and G.~P.~Dube,
Trans.~Faraday Soc.~\textbf{35}, 1125 (1939).

\bibitem{Larsen-Grier} A.~M.~Larsen and D.~G.~Grier, Nature
\textbf{385}, 230 (1997).

\bibitem{Kepler-Fraden} G.~M.~Kepler and S.~Fraden,
  Phys.~Rev.~Lett.~\textbf{73}, 356 (1994).

\bibitem{Carbajal-etal} M.~D.~Carbajal-Tinoco, F.~Castro-Rom\'an and
J.~L.~Arauz-Lara, Phys.~Rev.~E \textbf{53}, 3745 (1996).

\bibitem{Neu}
        J.C. Neu, Phys. Rev. Lett. {\bf 82}, 1072 (1999).
        
\bibitem{Sader}
        J.E. Sader and D.Y. Chan, 
        Langmuir {\bf 16}, 324 (2000).
        
\bibitem{Trizac-Raimbault-charge-like-steric} 
E.~Trizac and J.-L.~Raimbault,
Phys.~Rev.~E \textbf{60}, 6530 (1999). 

\bibitem{Trizac-charge-like} 
E.~Trizac, Phys. Rev. E \textbf{62}, R1465 (2000).

\bibitem{Brandes-Lue} T.~Brandes and L.~Lue, \textit{Fluctuations
  induced effects in confined electrolyte solutions}, e-print
  cond-mat/0311375v1, (2003).

\bibitem{Torres-Tellez-finite-size-DH} A.~Torres and G.~T\'ellez,
J.~Phys.~A: Math.~Gen.~\textbf{37}, 2121 (2004).

\bibitem{Janco-Tellez-ideal-conductor-limit}
B.~Jancovici and G.~T\'ellez, 
J.~Phys.~A: Math.~Gen.~\textbf{29}, 1155 (1996).


\bibitem{Aqua-Cornu-diel-wall1}
J.-N.~Aqua and F.~Cornu, J.~Stat.~Phys. \textbf{105}, 211 (2001).

\bibitem{Aqua-Cornu-diel-wall2}
J.-N.~Aqua and F.~Cornu, J.~Stat.~Phys. \textbf{105}, 245 (2001).

\bibitem{Aqua-these}
J.-N.~Aqua, \textit{Physique statistique des fluides coulombiens
  classiques et quantiques au voisinage d'une paroi}, PhD Thesis,
Universit\'e de Paris XI (2000).

\bibitem{LeboLieb-PRL}
J.~L.~Lebowitz and E.~Lieb, Phys.~Rev.~Lett.~\textbf{22}, 631 (1969).

\bibitem{LeboLieb-AdvMat}
E.~Lieb andJ.~L.~Lebowitz, Adv.~Math.~\textbf{9}, 316 (1972).

\bibitem{Alastuey-Janco-etc}
A.~Alastuey, B.~Jancovici, L.~Blum, P.~J.~Forrester, M.~L.~Rosinberg,
J.~Chem.~Phys.~\textbf{83}, 2366 (1985).

\bibitem{Kennedy} T.~Kennedy, 
Comm.~Math.~Phys. \textbf{92}, 269 (1983)

\bibitem{Torres-Tellez-general-DH} 
A.~Torres and G.~T\'ellez, 
\textit{General considerations on the finite-size corrections for
  Coulomb systems in the Debye--H\"uckel regime}, e-print
cond-mat/0404588 (2004).

\bibitem{Samuel} S.~Samuel, Phys.~Rev.~D \textbf{18}, 1916 (1978).

\bibitem{Caillol-loop} J.~M.~Caillol, 
J.~Stat.~Phys. \textbf{115}, 1483 (2004).

\bibitem{Lebo-Martin}
J.~L.~Lebowitz and Ph.~A.~Martin, 
J. Stat. Phys. \textbf{34}, 287 (1984).

\bibitem{Janco-screen-correl-revisit}
B.~Jancovici, J. Stat. Phys. \textbf{80}, 445 (1995)

\bibitem{GR}
S.~Gradshteyn and I.~M.~Ryzhik, \textit{Table of Integrals, Series, and
Products} (Academic, New York, 1965).

\bibitem{Samaj-Janco-TCP-metal} 
L.~\v{S}amaj and B.~Jancovici,
J.~Stat.~Phys. \textbf{103}, 717 (2001).

\bibitem{Janco-Tellez-coulcrit} 
B.~Jancovici and G.~T\'{e}llez,
J.~Stat.~Phys. \textbf{82}, 609 (1996).

\bibitem{AttardMitchellNiham-JCP}
P.~Attard, D.~J.~Mitchell, and B.~W.~Ninham,
J.~Chem.~Phys.~\textbf{88}, 4978 (1988).

\bibitem{Dean-Horgan-two-loop}
D.~S.~Dean and R.~R.~Horgan,
\textit{The two loop calculation of the disjoining pressure of a
  symmetric electrolyte soap film},
e-print cond-mat/0403150 (2004).

\bibitem{MerchanTellez-jabon-anillos-tcp}
L.~Merch\'an and G.~T\'ellez,
J.~Stat.~Phys. \textbf{114}, 735 (2004).

\bibitem{Lifshitz}
E.~M.~Lifshitz, Sov.~Phys.~JETP \textbf{2}, 73 (1956).


\bibitem{Manning} G.~S.~Manning J.~Chem.~Phys. \textbf{51}, 924
(1969).

\bibitem{Alexander} S.~Alexander, P.~M.~ Chaikin, P.~Grant,
G.~J.~Morales and P.~Pincus, J.~Chem.~Phys. \textbf{80}, 5776 (1984).

\bibitem{TellezTrizac-PB-asym} G.~T\'ellez and E.~Trizac,
\textit{Non-linear screening of spherical and cylindrical colloids:
the case of 1:2 and 2:1 electrolytes}, Phys.~Rev.~E (to be published),
e-print cond-mat/0403265 (2004).

\bibitem{Grahame} D.~C.~Grahame, J.~Chem.~Phys. \textbf{21}, 1054
(1953).

\bibitem{Ulander} J.~Ulander, H.~Greberg, and R.~Kjellander,
J.~Chem.~Phys. \textbf{115}, 7144 (2001).

\bibitem{Caillol-low-fugas}
J.~M.~Caillol and J.~L.~Raimbault,
\textit{J. Stat. Phys.} \textbf{103}, 753 (2001)

\bibitem{Caillol-high-temp}
J.~L.~Raimbault and J.~M.~Caillol,
\textit{J. Stat. Phys.} \textbf{103}, 777 (2001)

\bibitem{Janco-Samaj-TCP-dielec}
B.~Jancovici and L.~\v{S}amaj, J.~Stat.~Phys.~\textbf{104}, 755
(2001).

\bibitem{Janco-plane-wall-II} 
B.~Jancovici, J.~Stat.~Phys.~\textbf{29}, 263 (1982).

\bibitem{Aqua-Cornu-dipolar-PRE}
J.-N.~Aqua and F.~Cornu, Phys.~Rev.~E \textbf{68},
026133 (2003).

\bibitem{Martin-sum-rules-review}
Ph.~A.~Martin, Rev.~Mod.~Phys.~\textbf{60},
1075 (1988).

\bibitem{Tellez-modes}
G.~T\'ellez, Phys.~Rev.~E \textbf{55}, 3400 (1997).

\bibitem{Forrester-Janco-Tellez} 
P.~J.~Forrester, B.~Jancovici and G.~T\'ellez,
J.~Stat.~Phys. \textbf{84}, 359 (1996).

\bibitem{Levin-charge-attract}
Y.~Levin, Physica A \textbf{265}, 432 (1999).


\end{thebibliography}
\end{document}